\documentclass[doublecol,linenumbers]{epl2}
\usepackage{amsmath}
\usepackage{amssymb}
\usepackage[mathscr]{eucal}

\usepackage{graphicx}
\usepackage{lipsum}

\usepackage{color}
\definecolor{dgreen}{rgb}{0,0.7,0}

\newcommand{\ee}{\mathrm{e}}

\newcommand{\cY}{\mathcal{Y}}

\newcommand{\cC}{\mathcal{C}_b}

\newcommand{\omegain}{\omega_{in}}
\newcommand{\omegatyp}{\overline{\omega}}

\newcommand{\omegast}{\omega^*}
\newcommand{\hst}{h^*}
\newcommand{\of}{f}

\newcommand{\bG}{G_{ss}}

\newcommand{\chist}{\chi^*}

\usepackage{xr}
\title{Exact correlations in a single file system with a driven tracer}

\author{A. Kundu\inst{1} \and J. Cividini\inst{2}}
\shortauthor{A. Kundu and J. Cividini}

\institute{                    
	\inst{1} International center for theoretical sciences, TIFR, Bangalore - 560012, India\\
	\inst{2} Department of Physics of Complex Systems, Weizmann Institute of Science, Rehovot 76100, Israel
}


\date{\today}

\abstract{
We study the effect of  a single driven tracer particle in 
a bath of other particles performing the random average process on an infinite line using a stochastic hydrodynamics approach. 
We consider arbitrary fixed as well as random initial conditions and compute the two-point correlations. For 
quenched uniform and annealed steady state initial conditions we show that in the large time $T$ limit the fluctuations and the correlations of the positions of the particles grow {\it subdiffusively} as $\sqrt{T}$ and have well defined scaling forms under proper rescaling of the labels. We compute the corresponding scaling functions exactly for these specific initial configurations and verify them numerically. We also consider a non translationally invariant initial condition with linearly increasing gaps where we show that the fluctuations and correlations grow {\it superdiffusively} as $T^{3/2}$ at large times.
}

\begin{document}

\maketitle

\noindent
The motion of non-overtaking particles in narrow channels is known as single-file diffusion.
In such one dimensional geometry the motion of any tagged particle (TP) is hemmed by its neighbors.  
The study of TP motion have been started independently by 
Harris and Jepsen \cite{Jepsen-65, Harris-65} in 1965. Since then there have been numerous theoretical \cite{Percus-74, Beijeren-83, Arratia-83, Alexander-78, Majumdar-91, Lizana-09, Barkai-09, Kollmann-03, Gupta-07, Rodenbeck-98,
	Barkai-10, Roy-13, Sabhapandit-07, Illien-13, Benichou-13, Krapivsky-14, Hegde-14} as well as experimental 
studies~\cite{Gupta-95, Kulka-96, 
	Hahn-96, Wei-00, Meersmann-00, Lin-05, Lutz-04}.
One interesting feature is that the variance of the TP grows 
subdiffusively ($ \sim \sqrt{T}$) even when individual particles are diffusive and the prefactor of $\sqrt{T}$ depends on the initial condition (IC)~\cite{Harris-65, Percus-74, Beijeren-83, Arratia-83, Alexander-78, Majumdar-91, Barkai-09, Kollmann-03, Gupta-07, Rodenbeck-98,
	Barkai-10, Roy-13, Illien-13, Benichou-13, Krapivsky-14, Hegde-14}. 
Due to recent developments of experimental techniques like active microrheology and microfluidics \cite{Wilson-11, Wittbracht-10} there has been a resurgence of interest in studying locally driven 
interacting particle systems.
In this paper we consider the case where the TP is driven externally and study its effect on the surrounding unbiased bath particles.

Single driven TPs (DTPs) in quiescent media have been 
used to probe rheological properties of complex media such as polymers \cite{Gusche-08, Krager-09}, granular media \cite{Candelier-10, Pesic-12} or colloidal crystals \cite{Dullens-11}, and also to study directed cellular motion in crowded channels \cite{Hawkins-09}
or the active transport of vesicles in a crowded axon \cite{Loverdo-08}.
Some practical examples of DTPs are a charged
impurity being driven by an applied electric field or a colloidal particle being pulled by an optical tweezer in the presence of other
colloidal particles performing random motion. 

On the theoretical side, situations have been considered where 
the surrounding medium is a Symmetric Simple Exclusion Process. 
In this context the effect of the DTP has been quantified in terms of both the tracer motion and 
the perturbation of the density profile~\cite{Burlatsky-92,Burlatsky-96,Coninck-97,Landim-98,Benichou-99} in the Steady State (SS). 
In particular it has been shown by using a decoupling approximation that both the mean~\cite{Burlatsky-92, Burlatsky-96, Landim-98} 
and the variance~\cite{Burlatsky-96} of the position of the DTP grow as $\sqrt{t}$.
The full distribution of the displacement of the DTP has
recently been computed in the high density limit of bath particles~\cite{Illien-13}. However, all these studies always rely on some approximation, and mostly considered quantities related to the displacement of the TP.

\begin{figure}[t]
	\begin{center}
		\includegraphics[width=0.45\textwidth]{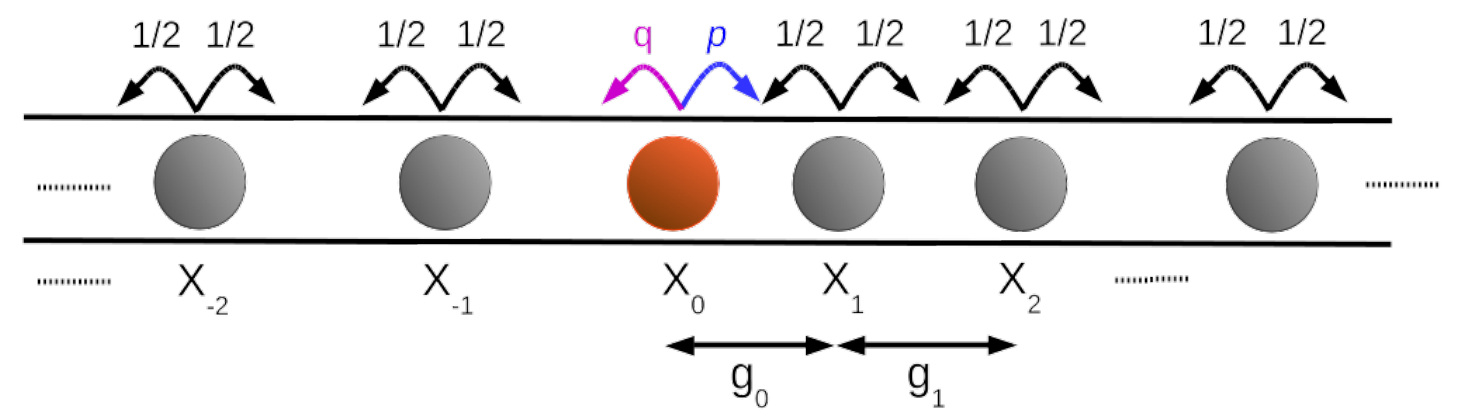} 
	\end{center}
	\caption{\small Schematic diagram of the RAP with a driven tracer particle 
		( $0$-th particle colored in red ) on an infinite line. $x_i$ represents the position of the 
		$i$-th particle and $g_i=x_{i+1}-x_{i}$ represents the gap between the $(i+1)$-th and $i$-th particles. The driven tracer particle hops to the left with 
		rate $p$ and to the right with rate $q$, whereas all other particles ($i\neq 0$) hop to the left or to the right with same rate $1/2$. }  
	\label{fig:scheme}
\end{figure}

Here we consider the motion of a DTP moving inside a pool of other particles performing the random average process (RAP) in one dimension. Using a stochastic hydrodynamic description for this model with a DTP, we compute 
the mean and the fluctuation of the position of not only the DTP but of \emph{all} the particles in the large time limit \emph{exactly}. 
In addition, we also compute the exact large time behavior of the correlation function between the positions of \emph{any} two particles.

The RAP was introduced in~\cite{Ferrari-98}. It appears in many problems like force propagation in granular media \cite{Coppersmith-96, Rajesh-00}, 
in the porous medium equation \cite{Feng-96}, 
in models of mass transport \cite{Krug-00, Rajesh-00},
voting systems \cite{Melzak-76} or wealth distribution \cite{Ispolatov-98} and in the generalised Hammersley process \cite{Aldous-95}.

\noindent
\emph{RAP Model}: 
We consider an infinite number of particles labeled by integers $k \in \mathbb{R}$ on the infinite line (see fig. \ref{fig:scheme}). We denote the position of the particle $k$ at time $t$ by 
$x_k(t)~\forall~k\in \mathbb{Z}$. 
Each particle is allowed to move to the right or left by a \textit{random fraction} $\eta \in [0,1)$
of the space available until the nearest particle in the direction of hopping. 
The fraction $\eta$ is drawn from some distribution $R(\eta)$ with moments $\mu_k =\int_0^1 d \eta~ \eta^k R(\eta)$. 
All particles hop towards right and left with rate $1/2$ except for the $0$-th particle which 
moves asymmetrically, to the right with rate $p$ and to the left with rate $q$. Without loss of generality we assume $p>q$.

The ICs of the particles, more precisely the initial separations $\omegain$ between successive particles are chosen from 
some arbitrary distribution $\mathcal{P}_{in}\asymp e^{-G[\omegain]}$. We are mainly interested in the following three ICs: 

\noindent
\textit{Quenched uniform (QU) initial conditions: }  At initial time the paticles are at fixed positions ('quenched') with a constant spacing ('uniform'). We therefore consider $x_k(0)=k \nu_0$, where $\nu_0^{-1}$ is the particle density in the RAP.
%For the quenched uniform initial conditions we consider $x_k(0)=k \nu_0$, where $\nu_0^{-1}$ is the particle density in the RAP. 
	
	\noindent
	\textit{Quenched linear (QL) initial conditions: }  For this case, the particles are arranged in such a way that the gaps between successive particles $g_k=x_{k+1}-x_{k}=\alpha~[\text{sgn}(k)~(k+1/2)+\delta_{k,0}]$ grows linearly with indices $k=\pm1,\pm2,...$ on both sides of the DTP ($k=0$). 

	\noindent
	\textit{Steady state (SS) initial conditions:} In the steady state initial condition, a configuration is chosen randomly from the steady state distribution 
	obtained in absence of DTP.
	This distribution is explicitly given in terms of the initial gap profile $\omega_{in}(z)$ by $\text{Prob.}[\omega_{in}] \asymp \exp (-\bG[\omega_{in}])$,
	\begin{eqnarray}
	\bG[\omega] = \int_{-\infty}^\infty dz~ [f(\omega)-f(\nu_0) - f'(\nu_0)(\omega-\nu_0)]~~~
	\end{eqnarray}
	with $f(\omega)=(1-\mu_1/\mu_2)\ln (\omega)$, $f'(\omega)=df/d\omega$. This expression of $\bG[\omega]$ can be obtained by following the procedure detailed in~\cite{Bertini-02,Bertini-15}, which mainly consists in deriving an appropriate hydrodynamic description starting from a microscopic point of view. 
	
The SS IC is the one expected to be realized in most experiments~\cite{Gusche-08, Candelier-10, Pesic-12, Dullens-11}, as the tracer particle is usually held and the rest of the fluid left untouched. Taking mesurements for the QU IC would require holding all the particles initially at fixed positions using either optical/magnetic tweezers or a periodic field, and then turning them off before starting the noisy dynamics.
 We here mainly focus on the difference between QU and SS ICs. The QL IC case is briefly discussed to demonstrate the possibility of superdiffusive behaviour.

\noindent
\emph{Results}: 
In the large time limit, we compute the mean, variance and correlations of the particles for {\it arbitrary} ICs. In particular, we show that for both QU and SS ICs, 
the mean $y_k(T)=\langle x_k(T)\rangle$ and the two point correlation $c_{kl}(T)=\langle x_k(T)x_l(T)\rangle-\langle x_k(T)\rangle \langle x_l(T)\rangle$ 
have the following scaling forms 
\begin{eqnarray}
	y_k(T)&=&\nu_0\sqrt{2\mu_1T}~\mathcal{Y}_b(k/\sqrt{2\mu_1T}), \nonumber \\ 
	c_{kl}(T)&=&\nu_0^{2}\sqrt{2\mu_1T}~\mathcal{C}_b(k/\sqrt{2\mu_1T},l/\sqrt{2\mu_1T}) \label{sclng-forms}
\end{eqnarray}
where $b=\frac{p-q}{p+q}$ is the bias and $\langle..\rangle$ denotes average over stochastic evolution as well as over ICs. 
These scaling behaviors \eqref{sclng-forms} have recently been observed in numerics and $\mathcal{Y}_b(x)$ has been obtained in~\cite{Julien_arxiv-16} 
using a microscopic point of view. Here we rederive 
it using a hydrodynamic approach and show that it is given by 
\begin{equation}
	\mathcal{Y}_b(u)= u +  b~[{e^{-u^2}}/{\sqrt{\pi}}-|u|\text{Erfc}(|u|)], \label{mcal-Y}
\end{equation}
for both QU and SS ICs.

For the zero drive case ($b=0$) the scaling function $\mathcal{C}_0(u,v)$ has been computed in~\cite{Rajesh-01} for both ICs. In this translationally invariant case the correlation function $\mathcal{C}_0(u,v)$ depends only on the separation $|u-v|$ between the two particles. On the other hand, in presence of DTP the system is not translationally invariant and the correlation functions depend on the scaling variables $u$ and $v$ individually.

For the driven case an attempt to compute the scaling function $\mathcal{C}_b(u,v)$ for the QU IC has been made 
but for small drive strength $|p-q|$ and for the totally asymmetric choice $q=0$ (or $b=1$) \cite{Julien_arxiv-16}.
In the latter case $\mathcal{C}_1(u,v)$ has been computed only for $u\geq 0$ and $v\geq 0$~\cite{Julien_arxiv-16}.
Here we compute these scaling functions over the full 2D plane exactly for arbitrary $p$ and $q$ and for arbitrary ICs. In particular, we show that the variance of the position of the DTP 
is given by 
\begin{equation}
	c_{0,0}(T) = \nu_0^{2}\sqrt{2\mu_1T} ~\frac{\mu_2\sqrt{2}~ (\sqrt{2}-1)^2}{\sqrt{\pi}(\mu_1-\mu_2)} \mathcal{A}(b), \label{fluc-dtp}\\
\end{equation}
where $\mathcal{A}$ is given by
\begin{equation}
	\label{A_a}
	\mathcal{A}(b)= \begin{cases}
		& (\sqrt{2}+1)(1-b^2) + \frac{1}{2}(1+b^2),~~~~~~\text{QU} \\
		& (\sqrt{2}+1)(1-b^2) + \frac{2+\sqrt{2}}{2}(1+b^2),~~\text{SS}.
	\end{cases}
\end{equation}
The equations~\eqref{fluc-dtp},
\eqref{A_a} match with previously known results for $b=0$ for both ICs~\cite{Rajesh-01}, and $b=1$ in the QU case~\cite{Julien_arxiv-16}. Note that the ratio of the prefactors in the QU and SS ICs depends on the drive and becomes $\sqrt{2}$ for $b=0$, as observed earlier~\cite{Rajesh-01, Barkai-09, Krapivsky-14}.

\begin{figure}[h]
	\begin{center} 
		\includegraphics[scale=0.34]{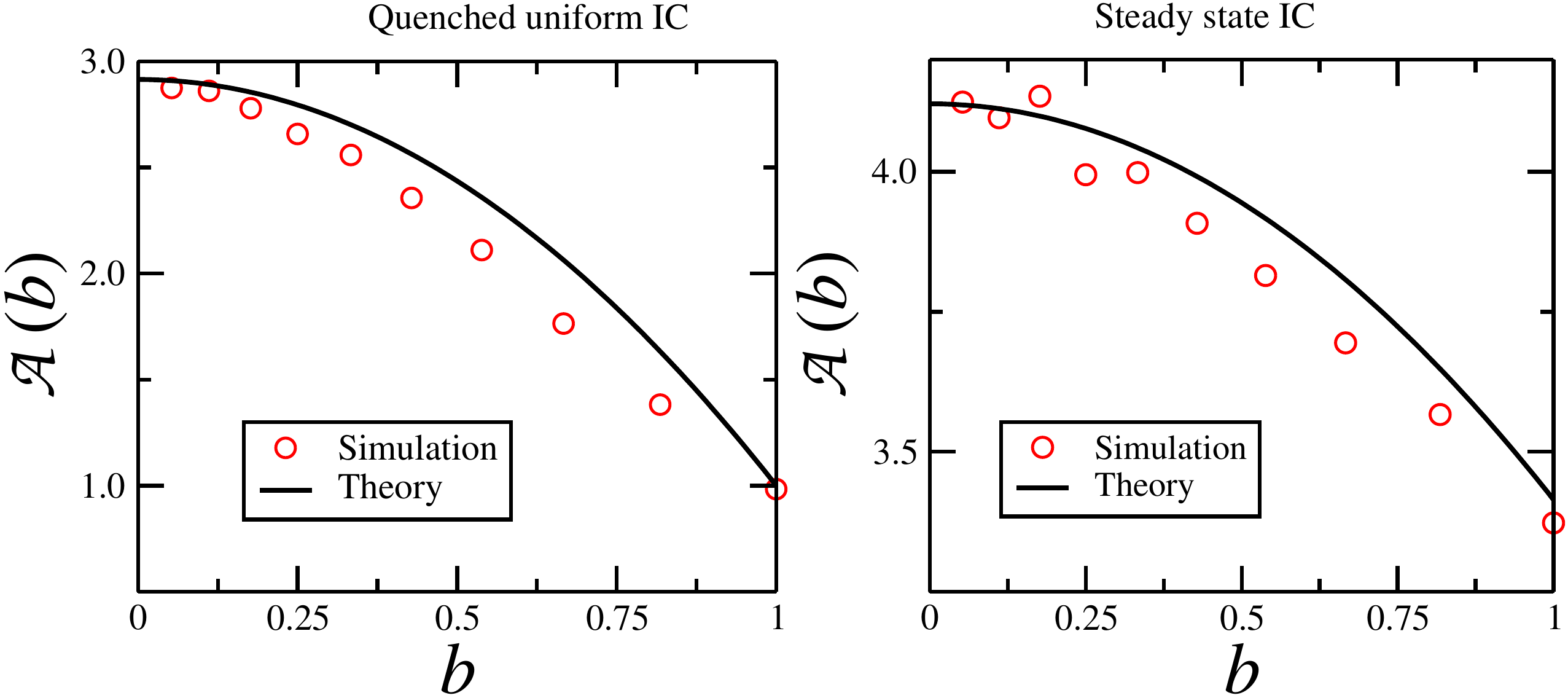} 
	\end{center}
	\caption{\small Numerical measurements of the variance of the DTP 
		(red circles) as function of $b$ for fixed $p=1$ compared with the theory~\eqref{fluc-dtp}-\eqref{A_a} (solid lines) for QU (left column) and SS (right column) ICs. Simulations are performed on a ring with $N=400$ particles.}  
	\label{fig:A_a}
\end{figure}

To prove the above results, we map the RAP to a mass transfer problem~\cite{Julien_arxiv-16}. The mass at site $k$ is equal to the 
the gap $g_k(t)=x_{k+1}(t)-x_k(t)$ between the positions of the $(k+1)$th and $k$th particles in RAP for all $k$.
The DTP is mapped to a special link $(-1,0)$
where mass transfer occurs with rate $p$ to the left and $q$ to the right while all the other links are symmetric.  
In terms of the gaps, the displacement of the $k$-th particle at time $T$
is given by  
$x_k(T)= \sum_{l=-\infty}^{k-1} [g_l(T)-g_l(0)]$.

In the large length and time scales it seems convenient to consider a continuum description of the system. For that we coarse-grain the gap label $k$ to get a continuous variable $z$ as done in~\cite{Gupta-07}. This allows us to describe the system in terms of  
the conserved mass density field $\omega(z,t)$ and the current field $j(z,t)$, which are related via the continuity equation $\partial_t \omega(z,t) = -\partial_z j(z,t)$.
The displacement of the $k$th particle can be expressed as 
\begin{equation}
	X_k(T)=x_k(T)-x_k(0)= \int_{-\infty}^k dz~[\omega(z,T)-\omega(z,0)]. \label{X_l-def} 
\end{equation}
Since the TP is driven, the region at its front will be more crowded than the region at its back, thus creating a lower gap density 
in front of it than at its back. This leads to the following discontinuity~\cite{Julien_arxiv-16}
\begin{equation}
	q ~\omega(z,t)|_{z\to 0^-}=p~\omega(z,t)|_{z\to 0^+}~. \label{BC-density-discontinuity} 
\end{equation}
Since the mass transfer model falls in the class of `gradient type' models, the typical (and also the average) gap density profile $\omegatyp(z,t)$ satisfies the diffusion equation~\cite{Kipnis-99, Eyink-90}
\begin{equation}
\partial_t \omegatyp = \partial_z [D(\omegatyp) \partial_z \omegatyp], \label{anyic} 
\end{equation}
where $D(\omega)=\mu_1/2$ is the diffusivity~\cite{Krug-00}. 
To solve \eqref{anyic} we introduce the Green's function $F_b(z,z',t)$ that satisfies $\partial_t F_b = \frac{\mu_1}{2} \partial_z^2 F_b$ with the discontinuity conditions $(1-b) F_b|_{z=0^-} = (1+b) F_b |_{z=0^+}$ and $\left(F_b\right)_{z=0^-}^{0^+} = 0$ 
	and the delta initial condition $F_b(z,z',0) = \delta (z-z')$. The Green's function $F_b$ can be determined by standard methods and is given by 
	\begin{equation}
	F_b(z,z',t) = \frac{e^{-\frac{(z-z')^2}{2 \mu_1 t}}}{\sqrt{2 \pi \mu_1 t}} - b~\text{sgn} (z) \frac{e^{-\frac{(|z|+|z'|)^2}{2 \mu_1 t}}}{\sqrt{2 \pi \mu_1 t}}, \label{Fa} 
	\end{equation}
One can solve for $\omegatyp(z,t)$ satisfying \eqref{BC-density-discontinuity} for arbitrary fixed IC $\omegain(z')$ to get~\cite{SM}
\begin{eqnarray}
	&&\omegatyp(z,t) = \int_{-\infty}^\infty F_b(z,z',t) ~\omegain(z')~ d z', \label{omega_0}
\end{eqnarray}
where $\text{sgn}(x)$ is the sign function.
Using $\langle \omegain(z')\rangle_{\mathcal{P}_{in}}=\nu_0$ 
both for QU and SS ICs, in the above equation we explicitly get 
$\omegatyp(z,T)=\Omega_b(z/\sqrt{2\mu_1T})$ with $\Omega_b(u) = \nu_0[1 -b~\text{sgn}(u)~\text{erfc}(|u|)]$. Inserting this result in~\eqref{X_l-def} we obtain~\eqref{sclng-forms}-\eqref{mcal-Y}. 

For the QL IC case,
where the particles are initially placed at positions $x_k(0)=(\alpha/2)~\text{sgn}(k)~k^2,~\forall k \in \mathbb{Z}$ one has $\omegain(z')= \alpha |z'|$. Using this in \eqref{omega_0} and performing the integral  one gets $\omegatyp(z,T) = \sqrt{2\mu_1T}~\alpha~\Omega_b^{QL}(z/\sqrt{2\mu_1T})$ where 
$\Omega_b^{QL}(u)=|u|+[1-b~\text{sgn}(u)][{e^{-u^2}}/{\sqrt{\pi}}-|u|\text{Erfc}(|u|)]$. Once again inserting this result in \eqref{X_l-def} one obtains $y_k(T) \sim T$ in contrast to $\sim \sqrt{T}$ behaviour: $y_k(T)=2\mu_1T~\alpha~\cY_b(k/\sqrt{2\mu_1T})$ where 
$\cY_b(u)=\text{sgn}(u)~u^2+\Theta(u)/2+[b-\text{sgn}(u)][(1+2u^2)\text{erfc}(|u|)-2|u|e^{-u^2}/\sqrt{\pi}]/4 $ and $\Theta(x)$ is the Heaviside theta function.

We now focus on computing the variances and the two point correlations. For that we need to consider the fluctuations of the density and current fields about their mean behaviors.
According to the macroscopic fluctuation theory~\cite{Bertini-15, Krapivsky-14}, the fluctuations in these two fields can be described by adding Gaussian fluctuations to the current:  
$j(z,t)= -(\mu_1/2)\partial_z \omega(z,t) + \sqrt{\sigma(\omega(z,t))}\eta(z,t)$ where $\sigma(\omega)$ is 
the mobility and $\eta(z,t)$ is a Gaussian white noise satisfying $\langle \eta(z,t) \rangle=0$ and $\langle \eta(z,t) \eta(z',t')\rangle=\delta(z-z')\delta(t-t')$. For 
the mass transfer model one can show $\sigma(\omega)=\frac{\mu_1\mu_2}{\mu_1-\mu_2}\omega^2$~\cite{Krug-00}. 
Due to the Gaussian nature of the noise, the joint probability of observing density and current profiles 
$\omega(z,t)$ and $j(z,t)$ different from the typical profiles is also Gaussian and is given by~\cite{Bertini-15, Krapivsky-14}
\begin{eqnarray}
	&&\mathbb{P}[\omega,j] \asymp \exp\left[-\int_0^T dt \int_{-\infty}^\infty dz \frac{\left[ j+ (\mu_1/2)\partial_z \omega(z,t) \right]^2}{2 \sigma(\omega(z,t))^2}\right]~~~ \label{P-rho-j}
\end{eqnarray}
where $(\omega,j)$ satisfy~the continuity equation. Here the symbol $\asymp$ represents logarithm equivalence.

In order to compute $c_{k,l}(T)$ we look at the joint distribution of 
the displacements $X_k(t)$ and $X_l(t)$. This distribution is 
completely characterized by the 
generating function
\begin{equation}
	\mu (\lambda_k,\lambda_l) = \langle \exp \{\lambda_kX_k(T)+\lambda_lX_l(T)\} \rangle, \label{cumu-func}
\end{equation}
where $\lambda_{k,l}$ are Lagrange multipliers. 
From this quantity the two-particle connected correlations can be obtained as  
$c_{k,l}(T)=[{\partial^2 \ln [\mu(\lambda_k,\lambda_l)]}/{\partial \lambda_k \partial \lambda_l}]_{\lambda_k=0,\lambda_l=0}$.

To proceed further we follow the technique developed by Krapivsky \emph{et al.}~\cite{Krapivsky-14} and 
compute $\mu (\lambda_k,\lambda_l)$ using the joint distribution~\eqref{P-rho-j}. We consider an ensemble 
of ICs $\omegain$ drawn with probability $\mathcal{P}_{in}\asymp e^{-G[\omegain]}$. 
From \eqref{P-rho-j} and~\eqref{cumu-func} we see that $\mu (\lambda_k,\lambda_l)$
can be expressed as a path integral 
\begin{equation}
	\mu (\lambda_k,\lambda_l)= \int \mathcal{D}[\omegain]  e^{-G[\omegain]}\int_{\omega|_{t=0} =\omegain} \hspace{-12mm} \mathcal{D}[\omega,h]e^{-S[\omega,h]}, \label{Path-int}
\end{equation}
where $h$ is an auxiliary field. We here stress the fact that the integrals over $\omega$ and $h$ have to be evaluated for a fixed IC $\omegain$ before taking the average over initial configurations. The action $S$ is obtained via Martin-Siggia-Rose formalism~\cite{Krapivsky-14},
\begin{eqnarray}
	&&S[\omega,h] = -\lambda_kX_k(T) - \lambda_l X_l(T) + \int_0^T \int_{-\infty}^\infty dz \left [ h~\partial_t \omega \right.   \nonumber \\
	&&~~~~~~~~~~~~\left. +(\mu_1/2)(\partial_z\omega)(\partial_z h) - ({\sigma(\omega)}/{2})(\partial_z h)^2\right],~~ \label{action}  
\end{eqnarray}
where we note that the integrand is singular at $z=0$ at all times.
At large times $T$ the path integral over $\omega$ and $H$ can be computed by saddle point method to give $\sim \ee^{-S[\omegast[\omegain],\hst[\omegain]]}$, as it is dominated by optimal paths $(\omegast(z,t),\hst(z,t))$ that depend on the initial configuration $\omegain$. In the final step one has to perform the remaining path integral over $\omegain$ in \eqref{Path-int} in the annealed case. This can again be carried out by saddle-point method.

We now have to find the optimal paths $(\omegast(z,t),\hst(z,t))$ for arbitrary fixed IC $\omegain(z)$. Using variational calculus we obtain that the optimal paths satisfy the following coupled differential equations~\cite{SM}
\begin{eqnarray}
	&&\partial_t \omegast(z,t)= (\mu_1/2)\partial_z^2 \omegast - \partial_z(\sigma(\omegast)~\partial_z \hst), \label{opt-diff-eq-1} \\
	&&\partial_t \hst(z,t)= -(\mu_1/2)\partial_z^2 \hst - (\sigma'(\omegast)/2)~(\partial_z \hst)^2, ~~\label{opt-diff-eq-2}
\end{eqnarray}
where $\sigma'(\omega)=d\sigma /d\omega$. Writing $A(z)|_{0^+}^{0^-}=A(0^-)-A(0^+)$ for short, one can show that the solutions of \eqref{opt-diff-eq-1} and \eqref{opt-diff-eq-2} have to satisfy the discontinuity conditions
\begin{eqnarray}
	&&q ~\omegast(z,t)|_{z\to 0^-}=p~\omegast(z,t)|_{z\to 0^+}~,~\hst(z,t)|_{0^+}^{0^-}= 0 \nonumber \\
	&&\frac{\mu_1}{2}\left[\partial_z \omegast \right]_{0^+}^{0^-}=
	\left[\sigma(\omegast)\partial_z \hst\right]_{0^+}^{0^-},~\left[\rho\partial_z \hst\right]_{0^+}^{0^-}=0\label{discontinuities} 
\end{eqnarray}
and the initial/boundary conditions :
\begin{eqnarray}
	\omegast(z,0) &=& \omegain(z),~~\omegast(z,t)|_{|z|\to \infty}= \omegatyp(z,t)_{|z|\to \infty}, \nonumber \\ 
	\hst(z,T)&=& \lambda_k \Theta(k-z)+\lambda_l \Theta(l-z). \label{f-BCs} 
\end{eqnarray}
Hydrodynamic equations similar to \eqref{opt-diff-eq-1}-\eqref{opt-diff-eq-2} also appear in different other single file systems like Brownian motion models, exclusion processes \cite{Krapivsky-14}. These equations are normally {\it non-linear} and are usually hard to solve.  In presence of a DTP, additional complications appear as the solutions generically become singular at the tracer position. However it turns out that for RAP these hydrodynamic equations can be solved perturbatively as we show in the following.

To obtain the two point correlations one only needs to know the cumulant generating function $\ln \mu(\lambda_k,\lambda_l)$  up to $\mathcal{O}(\lambda^2)$. Hence it is enough to solve \eqref{opt-diff-eq-1}-\eqref{opt-diff-eq-2} to $\mathcal{O}(\lambda)$. 
We therefore expand the paths in powers of $\lambda_k$ and $\lambda_l$,
\begin{eqnarray}
\omegast& =& \omegatyp + \lambda_k\omega_k+\lambda_l\omega_l + \mathcal{O}(\lambda^2), \label{expnsn-1} \\
\hst&=& \lambda_kh_k+\lambda_lh_l + \mathcal{O}(\lambda^2).\label{expnsn-2}
\end{eqnarray}
Inserting these expansions in \eqref{opt-diff-eq-1}-\eqref{opt-diff-eq-2} and equating terms of order $\lambda_k$ and $\lambda_l$ from both sides separately, we obtain equations for the fields $\omega_\epsilon$ and $h_\epsilon$~\cite{SM}, 
\begin{eqnarray}
\partial_t \omega_\epsilon(z,t)&=&\frac{\mu_1}{2} \partial_z^2 \omega_\epsilon -\partial_z [\sigma(\omegatyp)~\partial_z h_\epsilon], \label{1st-ordr-opt-diff-eq-1} \\
\partial_t h_\epsilon(z,t)&=& -\frac{\mu_1}{2}\partial_z^2 h_\epsilon , \label{1st-ordr-opt-diff-eq-2}
\end{eqnarray}
for $\epsilon=(k,l)$ with BCs   
\begin{eqnarray}
\omega_{\epsilon}(z,0) &=& 0,~\omega_{\epsilon}(z,t)|_{|z|\to \infty} \to 0, \label{phi_bc} \\
\text{and}~h_\epsilon(z,T)& =& \Theta(\epsilon-z) \label{h_bc}
\end{eqnarray}
obtained from \eqref{f-BCs}. The discontinuities  
are obtained from~\eqref{discontinuities} as :
\begin{eqnarray}
&&(1+b) \omega_\epsilon(0^+,t)=(1-b)~\omega_\epsilon(0^-,t)~,\nonumber \\
&&
\frac{\mu_1}{2}\left[\partial_z \omega_\epsilon \right]_{0^+}^{0^-}=
\left[\sigma(\omegatyp)\partial_z h_\epsilon\right]_{0^+}^{0^-},~~~~ \label{dis_phi_2} \\
&&h_\epsilon(z,t)|_{0^+}^{0^-}= 0,\nonumber\\&& (1+b) \partial_z  h_\epsilon|_{0^-}=(1-b)\partial_z h_\epsilon|_{0^+},\label{dis_h} 
\end{eqnarray}
where $b=\frac{p-q}{p+q}$.
We first solve \eqref{1st-ordr-opt-diff-eq-2} with conditions \eqref{h_bc} and \eqref{dis_h}. Using a similar procedure as for the density profile, one can show that the solutions are given by 
	\begingroup\makeatletter\def\f@size{8}\check@mathfonts
	\def\maketag@@@#1{\hbox{\m@th\large\normalfont#1}}%
	\begin{eqnarray}
	h_\epsilon(z,t) &=& \frac{1}{2} \text{erfc} \left( \frac{z-\epsilon}{\sqrt{2 \mu_1 (T-t)}} \right) 
	+\frac{b}{2} \text{erfc} \left( \frac{|z|+|\epsilon|}{\sqrt{2 \mu_1 (T-t)}} \right).~~~~~ \label{sol-h_k}
	\end{eqnarray}
	\endgroup 
Now that we have $h_\epsilon(z,t)$ explicitly, we can solve \eqref{1st-ordr-opt-diff-eq-1}. The solution is given by 
\begin{eqnarray}
\omega_\epsilon(z,t)&=& -\int_0^td\tau \int_{-\infty}^\infty dz' F^b(z,z',t-\tau)  \\ && \times \partial_{z'} [\sigma(\omegatyp(z',\tau))~\partial_{z'} h_\epsilon(z',\tau)],\nonumber \label{sol-omega_k}
\end{eqnarray}
	$F^b(z,z',t)$ is given in \eqref{Fa}.
	The action in \eqref{action} depends on the integrals like $\Phi_\epsilon(z,t)=\int_{-\infty}^z\omega_\epsilon(z,t)dz$. Using 
	$F^b(z,z',t)= \partial_zh_z(z',T-t)$ and performing some manipulations, we obtain
	\begingroup\makeatletter\def\f@size{9}\check@mathfonts
	\def\maketag@@@#1{\hbox{\m@th\large\normalfont#1}}%
	\begin{equation}
	\Phi_\epsilon(z,t) = \int_0^td\tau \int_{-\infty}^\infty dz'~\sigma(\omegatyp(z',\tau))~\partial_{z'} h_\epsilon(z',\tau)~\partial_{z'} h_z(z',\tau), \label{sol-Phi} \\
	\end{equation}
	\endgroup
We can now insert $h_\epsilon$ and $\Phi_\epsilon$ functions in \eqref{action} and simplify using eqs.~\eqref{1st-ordr-opt-diff-eq-1}-\eqref{1st-ordr-opt-diff-eq-2} and find
\begin{eqnarray}
	&&S[\omegast,\hst] = -\frac{1}{2}\int_0^Td\tau \int_{-\infty}^\infty dz ~\sigma(\omegatyp)~(\lambda_k \of_k^b + \lambda_l \of_l^b)^2  \nonumber \\
	&& -\lambda_k \int_{-\infty}^k dz (\omegatyp-\omegain)-\lambda_l \int_{-\infty}^l dz (\omegatyp-\omegain) + \mathcal{O}(\lambda^2), ~~~~~\label{action-1}
\end{eqnarray}
where $\omegatyp(z,t)$ and $\of_\epsilon^b(z,t) = F_b(-\epsilon,-z,T-t)$ are given in \eqref{omega_0} and \eqref{Fa}. 
Once we have the optimal value of the action $S$ for given IC $\omegain$, we have to perform the average over the initial configurations by evaluating the remaining path integral over $\omegain$ in \eqref{Path-int}.

\begin{figure}[h]
	\begin{center} 
		\includegraphics[scale=0.34]{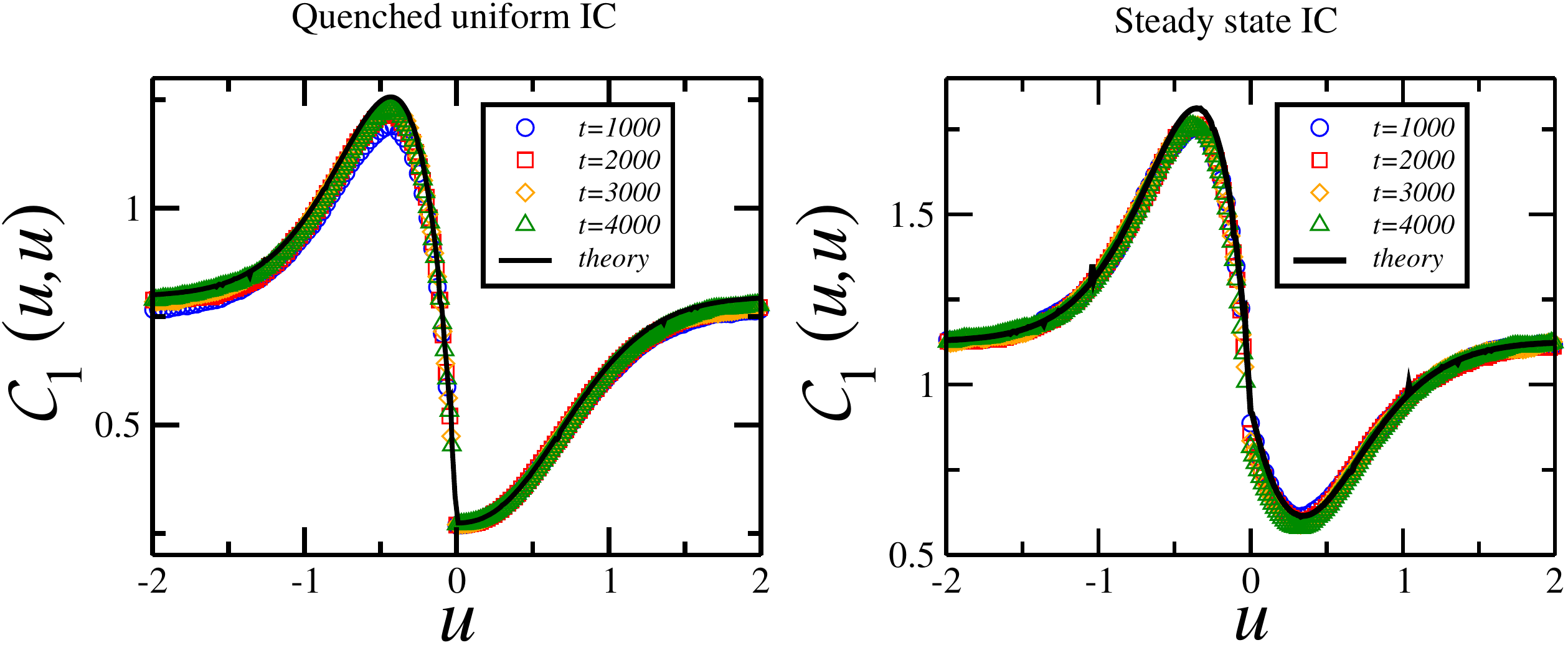} 
	\end{center}
	\caption{\small Numerical verification of the scaling behavior of $c_{k,k}(T)$ in \eqref{sclng-forms} and the corresponding 
		scaling function $\mathcal{C}_b(u,u)$ for $b=1$ for QU (left column) and SS (right column) ICs, and for $N=400$ as in fig.\,\ref{fig:A_a}. The symbols represent numerical data for increasing times while the black solid 
		line are the theoretical curves~\eqref{Cqu} and~\eqref{Css}. }  
	\label{fig:C_1-uu}
\end{figure}

\begin{figure}[h]
	\begin{center}
		\includegraphics[scale=0.34]{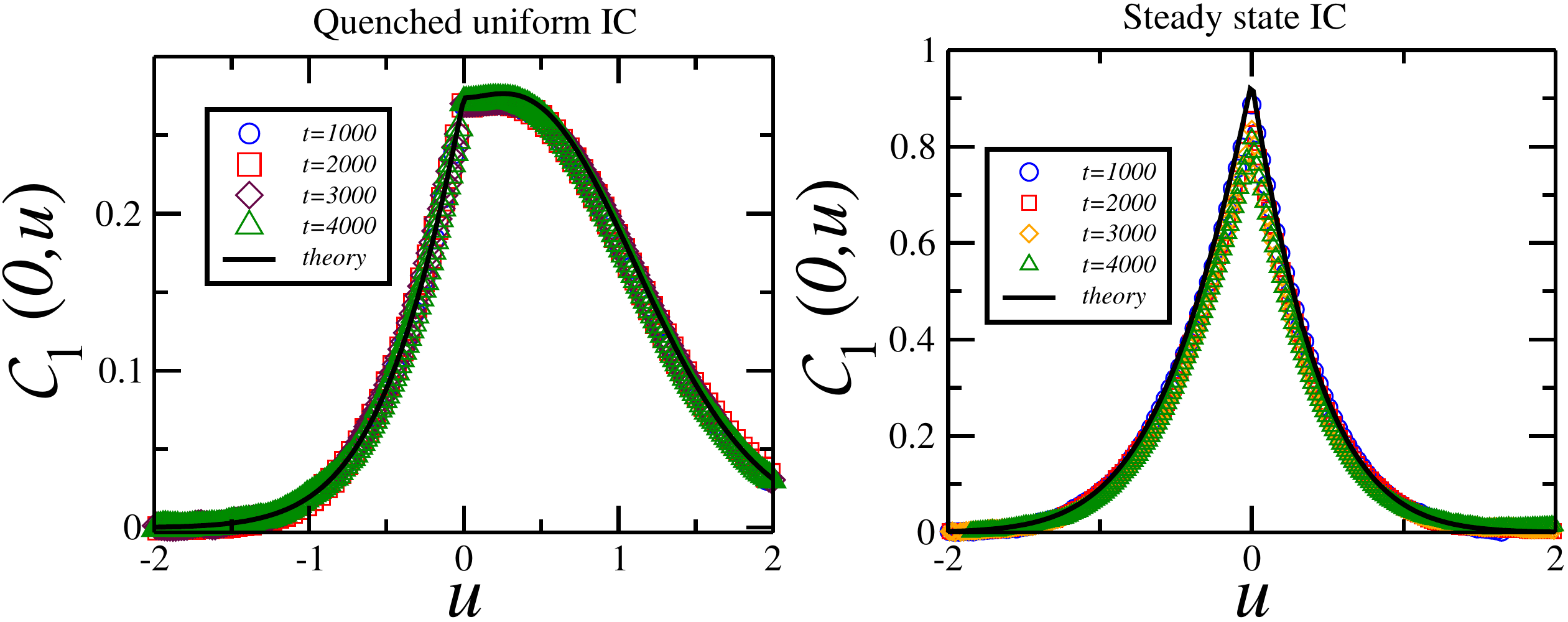} 
	\end{center}
	\caption{\small Numerical verification of $\mathcal{C}_1(0,u)$. The parameters are the same as in fig.\,\ref{fig:C_1-uu}.}  
	\label{fig:C_1-0u}
\end{figure}

To include all the ICs SS , QU and QL in a single analysis, we consider a general Gaussian distribution of $\omegain$, $G[\omegain] \simeq \int dz ~\frac{(\omegain(z)-\chi(z))^2}{2 \Sigma(z)^2}$.
The SS, QU and QL ICs are obtained by taking  $(\chi,\Sigma^2) \equiv (\nu_0,\frac{\mu_2 \nu_0^2}{\mu_1-\mu_2})$, $ (\nu_0,0)$ and $ (\alpha |z|,0)$, respectively. 
In \cite{SM} we show that 
\begin{eqnarray}
	&&c_{k,l}[\chi,\Sigma]= \int_0^Td\tau \int_{-\infty}^\infty dz~ \sigma(\omegatyp[\chi](z,t)) \of_k^b(z,t)\of_l^b(z,t) \nonumber  \\ &&
	~~~~~~~~~~~+ \int_{-\infty}^\infty dz~
	\Sigma(z)^{2} \mathcal{F}_k^b(z,T)\mathcal{F}_l^b(z,T),~\text{with}\label{C-gaussian}
	\\
	&& 
	\mathcal{F}_k^b(z,T)= 
	\frac{1}{2}\left(\text{erfc}\left(\frac{|u-\epsilon|}{\sqrt{2 \mu_1  t}}\right) + b~ \text{erfc}\left(\frac{\left| \epsilon\right| +\left| u\right| }{\sqrt{2 \mu_1  t}}\right)\right).\nonumber
\end{eqnarray}
The notation $\omegatyp[\chi](z,t)$ emphasizes that the average density profile $\omegatyp$ depends on $\chi$ but not on $\Sigma$ as can be seen from \eqref{omega_0}. The first term comes from the time evolution only, while the second term, proportional to $\Sigma^{2}$, comes from the variance of the initial configuration. In particular this expression shows that the variance $c_{k,k}[\chi,0] < c_{k,k}[\chi,\Sigma]$ for any choice of $\chi$ and $\Sigma$ and for any particle $k$. 

In general the integrals appearing in~\eqref{C-gaussian} can easily be evaluated numerically for arbitrary initial conditions after inserting the corresponding $\chi$ and $\Sigma^2$. In some special cases further analytical simplifications are possible. For the QU and SS ICs the integrals in $c_{0,0}$ can be performed analytically and one can obtain the variance of the DTP as given in \eqref{fluc-dtp} and \eqref{A_a}. In fig.~\ref{fig:A_a} we compare these predictions with numerical simulations and find nice agreement. Similarly, for these two ICs in the totally asymmetric case $b=1$, one finds the following expressions of the scaling functions defined in~\eqref{sclng-forms} after inserting $(\chi,\Sigma^2) \equiv (\nu_0,\frac{\mu_2 \nu_0^2}{\mu_1-\mu_2})$ and $ (\nu_0,0)$ in \eqref{C-gaussian} for SS and QU ICs, respectively and simplifying. Writing $a=\frac{\mu_2}{2\pi(\mu_1-\mu_2)}$ for short, we get
\begingroup\makeatletter\def\f@size{8}\check@mathfonts
\def\maketag@@@#1{\hbox{\m@th\large\normalfont#1}}%
\begin{eqnarray}
	&&	\mathcal{C}_{1}^{QU}(u,v) =a~\int_0^1 \frac{d\tau}{\tau}\int_{-\infty}^\infty  dy 
	\left( 1-\text{sgn}(y) \text{erfc}\left(\frac{y}{\sqrt{1-\tau}}\right)\right)^2   \times  \label{Cqu}
	\\ 
	&&
	\left(e^{-\frac{(y-u)^2}{\tau}} + \text{sgn}(y) e^{-\frac{(|y|+|u|)^2}{\tau}} \right) \left(e^{-\frac{(y-v)^2}{\tau}} + \text{sgn}(y) e^{-\frac{(|y|+|v|)^2}{\tau}} \right), \nonumber \\
	&&	\mathcal{C}_1^{SS}(u,v) = \mathcal{C}_1^{QU}(u,v) + \frac{\pi a}{2} \int_{-\infty}^\infty dy~\left[ \text{erfc}\left(|y-u|\right) +  \text{erfc}\left(|u|+|y|\right)\right] \nonumber \\
	&&\times \left[ \text{erfc}\left(|y-v|\right) +  \text{erfc}\left(|v|+|y|\right)\right].	\label{Css}
\end{eqnarray}
\endgroup
For the $b=1$ case we compare numerical results for $\mathcal{C}_1(u,u)$ and $\mathcal{C}_1(0,u)$ with the theory \eqref{Cqu} and~\eqref{Css} in figs.~\ref{fig:C_1-uu} and~\ref{fig:C_1-0u} for both ICs. As expected, these expressions are coherent with the results found in~\cite{Rajesh-01} and \cite{Julien_arxiv-16} in the appropriate cases.

In fig.~\ref{fig:C_1-uu} we find that the variances 
of the positions of particles in the front of the DTP are smaller than at the back, because the particles in the front 
have less space than those at the back. 
This intuitively suggests that the variance should be maximum just behind the DTP. On the contrary, in fig.~\ref{fig:C_1-uu} we see that the maximum occurs at a finite distance $\mathcal{O}(\sqrt{T})$ behind the DTP. It seems like the driven motion of the TP creates a density perturbation moving forward which is accompanied by a corresponding vacancy perturbation moving backwards similar to what happens in Exclusion Processes~\cite{Gupta-07}. 

Figure~\ref{fig:C_1-0u} displays the plot of the correlation (at time $T$) between the DTP and $i$th particle as a function of the scaling variable $u=i/\sqrt{2\mu_1 T}$. We notice that it has a vanishing slope for $u \to 0^+$ in the QU IC case implying that the particles at the front of the DTP are equally correlated up to a distance $\mathcal{O}(\sqrt{T})$. This is another artifact of the crowding phenomenon at $u>0$. On the other hand for SS ICs there is an extra contribution from the initial correlations that dominates at large times.

It is worth noting that the $\sqrt{T}$ growth of the correlation holds only for the translationally invariant ICs like QU and SS.  This however does not hold for non translationally invariant ICs like the QL IC  for which one has $\chi(z) =\alpha |z|$ and $\Sigma = 0$. 
In this case the DTP gets enough space on both sides and as a result the correlation grows {\it superdiffusively} $c_{k,l}(T) = T^{3/2}~\alpha^2~\cC^{QL}(k/\sqrt{2\mu_1T},l/\sqrt{2\mu_1T})$ at large times (see \cite{SM} for the proof).

In this work we consider a single-file system, the RAP, in presence of a locally driven particle which drives the whole system out of equilibrium. The drive manifests itself into discontinuities of the hydrodynamic fields at the DTP position. For this model we obtain  the large-time two-particle correlation function for \textit{any} IC exactly. The two-point correlation functions grow as $\sqrt{T}$ for both the QU and SS ICs, similar to the annealed/quenched dichotomy in non-driven systems. This is in contrast with Exclusion Processes, where nonlinearities play an important role in the quenched case \cite{Beijeren-91}. We also have looked at another specially designed quenched linear initial condition where the two point correlation functions grow as $\sim T^{3/2}$ and have scaling forms under rescaling of particle labels by $T^{1/2}$.  These correlation functions  will serve as a benchmark in analyzing the trajectories of a driven tracer particle in traffic flow, in active microrheology \cite{Wilson-11} and microfluidics \cite{Wittbracht-10} experiments.  An extension of the method used here to a larger class of single file systems would be of great interest.

\vspace{5mm}

We acknowledge useful discussions with S.~N.~Majumdar and D.~Mukamel. The support of the Israel Science Foundation (ISF) is gratefully acknowledged. AK would like to acknowledge the hospitality of the Weizmann Institute of Science where part of the work was done while he was visiting.

%\newpage

\begin{onecolumn}
\section{Continuity of $h(z,t)$ across $z=0$}

\noindent
According to the Macroscopic Fluctuation Theory the joint probability of the density and current profiles $\omega(z,t)$ and $j(z,t)$ is given by
\begin{eqnarray}
&&\mathbb{P}[\omega(z,t),j(z,t)] \asymp e^{-\mathbb{F}[\omega(z,t),j(z,t)]},~~\mathrm{where}, \label{P-rho-j}\\
&& \mathbb{F}[\omega,j] = \int_0^T dt \int_{-\infty}^\infty dz \frac{\left( j+ (\mu_1/2)\partial_z \omega(z,t) \right)^2}{2 \sigma(\omega(z,t))^2}, \label{mbb-F} \\
&&\mathrm{with}~~~~\partial_t \omega(z,t) = -\partial_z j(z,t). \label{continuity-eq}
\end{eqnarray}
From \eqref{P-rho-j} the probability of a given density profile $\omega(z,t)$ can be obtained by optimizing $\mathbb{F}[\omega(z,t),j(z,t)]$
over all the current profiles $j(z,t)$ satisfying \eqref{continuity-eq} and it is given by $\mathscr{P}[\omega(z,t)] \asymp e^{-\mathscr{F}[\omega(z,t)]}$ where 
$\mathscr{F}[\omega(z,t)] = \min_{j(z,t)} \mathbb{F}[\omega(z,t),j(z,t)].$
From the explicit expression of $\mathbb{F}$ in \eqref{mbb-F}, the optimizing equation reads 
\begin{equation}
\int_{-\infty}^\infty dz \frac{\left( j^o(z,t)+ D(\omega(z,t))\partial_z \omega(z,t) \right)}{\sigma(\omega(z,t))} =0,
\end{equation}
whose solutions can be written as $j^o(z,t)=-D(\omega)\partial_z \omega + \sigma(\omega)~\partial_z h$ where $h(z,t)$ is an auxiliary function. 
Putting this optimized solution $j^o(z,t)$ in $\partial_t \omega = \partial_z [D(\omega)\partial_z \omega -\sigma(\omega)~\partial_z h ]$,
one finds that the auxiliary function is continuous across $z=0$ : 
\begin{equation}
h(z,t)|_{z \to 0^+}= h(z,t)|_{z \to 0^-}. \label{H-continuity}
\end{equation}

\section{Optimal field equations for a given fixed initial configuration $\omegain(z)$}
%\section{Derivation of the optimal field equations and the associated boundary conditions for a given fixed initial configuration $\omegain(z)$ }
\noindent
Here we derive the equations \eqref{letter-opt-diff-eq-1} and \eqref{letter-opt-diff-eq-2} of the main text. We start with the generating function 
as defined in \eqref{letter-cumu-func} of the main text, where we do not yet perform the average over the initial condition:
\begin{equation}
\langle \exp[\lambda_kX_k(T)+\lambda_lX_l(T)] \rangle_{\omegain} = \int_{\omega|_{t=0} =\omegain} \mathcal{D}[\omega,h]e^{-S[\omega,h]}, \label{sm:Path-int}
\end{equation}
where the action, obtained via Martin-Siggia-Rose formalism, is given by
\begin{eqnarray}
&&S[\omega,h] = -\lambda_kX_k(T) - \lambda_l X_l(T) + \int_0^T dt \int_{-\infty}^\infty dz \left( h\partial_t \omega \right.   
\left. +(\mu_1/2)(\partial_z\omega)(\partial_z h) - ({\sigma(\omega)}/{2})(\partial_z h)^2\right),~~ \label{sm:action}  
\end{eqnarray}
where 
\begin{equation}
X_k(T)=x_k(T)-x_k(0)= \int_{-\infty}^k dz~[\omega(z,T)-\omega(z,0)]. \label{sm:X_l-def} 
\end{equation}
For large $T$, the integral in \eqref{sm:Path-int} is dominated by the optimal paths $(\omegast (z,t),\hst(z,t))$ 
that minimize the action and it is given by $\sim \exp({-S[\omegast ,\hst]})$.
To find out the optimal paths we consider small variations $\phi$ and $\psi$ of $\omega$ and $h$ and compute the variation of the action, 
\begin{eqnarray}
\delta  S &=& S[\omega +\phi,~h+\psi] - S[\omega ,~h], \nonumber \\
&=& - \int_{-\infty}^\infty dz~\left(\lambda_k \frac{\delta  X_k}{\delta \omega (z,0)} 
+ \lambda_l~\frac{\delta  X_l}{\delta \omega (z,0)} +h(z,0)\right)~\phi(z,0) \nonumber \\
&&-  \int_{-\infty}^\infty dz~\left(\lambda_k \frac{\delta  X_k}{\delta \omega (z,T)} 
+ \lambda_l~\frac{\delta  X_l}{\delta \omega (z,T)}-h(z,T) \right)~\phi(z,T)  \\
&& + \int_0^Tdt \int_{-\infty}^\infty dz \left( \psi(z,t)\partial_t \omega (z,t) 
-  \phi(z,t) \partial_t h(z,t) \right) \nonumber \\
&& - \int_0^Tdt \int_{-\infty}^\infty dz \left(\phi(z,t) \sigma'(\omega (z,t))~(\partial_zh(z,t))^2 
+ 2 \sigma(\omega (z,t))~ \partial_zh(z,t)~\partial_z\psi(z,t)) \right) \nonumber \\
&& + \frac{\mu_1}{2}\int_0^Tdt \int_{-\infty}^\infty dz \left(\partial_z\omega (z,t)~\partial_z\psi(z,t)) 
+ \partial_zh(z,t)~\partial_z\phi(z,t))\right) \nonumber \\
&=& \int_{-\infty}^\infty dz~ \left(\lambda_k~\Theta(k-z) + \lambda_l~\Theta(l-z) - h(z,0) \right)~\phi(z,0) \nonumber \\
&& + \int_{-\infty}^\infty dz~ \left(-\lambda_k~\Theta(k-z) - \lambda_l~\Theta(l-z) - h(z,T) \right)~\phi(z,T) \nonumber \\
&& - \int_0^Tdt \int_{-\infty}^\infty dz \left( \partial_th + \frac{\mu_1}{2} \partial_z^2h 
+ \frac{1}{2} \sigma'(\omega )~(\partial_zh(z,t))^2  \right)~\phi(z,t) \nonumber \\
&& + \int_0^Tdt \int_{-\infty}^\infty dz  \left( \partial_t \omega  - \frac{\mu_1}{2} \partial_z^2 \omega  
+ \partial_z( \sigma(\omega )~\partial_zh(z,t)~) \right)~\psi(z,t) \nonumber \\
&& +\frac{\mu_1}{2} \int_0^Tdt \left( \phi(z,t)~\partial_zh(z,t)~\right)_{0^+}^{0^-} 
+ \int_0^Tdt \left(\frac{\mu_1}{2} \psi(z,t)~\partial_z \omega (z,t) 
- \sigma(\omega ) \psi(z,t) \partial_zh(z,t) ~\right)_{0^+}^{0^-},  \nonumber 
\end{eqnarray}
where we used~\eqref{H-continuity} and for any observable $A$, $A|_{0^+}^{0^-}=A|_{z\to 0^+}-A|_{z\to 0^-}$.
Here $\phi(z,0)=0$ as we start from a fixed initial condition $\omega_{in}(z)$. 
The action is extremal for $(\omega,h) = (\omegast, \hst)$, implying $\delta  S = 0$. The optimal paths therefore satisfy 
\begin{eqnarray}
\partial_t \omegast  &=& \frac{\mu_1}{2} \partial_z^2 \omegast  - \partial_z[ \sigma(\omegast )~\partial_z\hst(z,t)~] \label{sm:opt-1} \\
\partial_t\hst &=& - \frac{\mu_1}{2} \partial_z^2\hst  - \frac{1}{2} \sigma'(\omegast )~[\partial_z\hst(z,t)]^2,\label{sm:opt-2}
\end{eqnarray}
with initial/final conditions 
\begin{equation}
\boxed{  \phi(z,0)=0 \implies \omegast(z,0)=\omegain(z),~~~\text{and}~~~\hst(z,T) = \lambda_k \Theta(k-z) + \lambda_l \Theta(l-z)} \label{sm:IC-1}
\end{equation}
% 
%  \begin{eqnarray}
% %   h(z,0) &=& \lambda_k \Theta(k-z) + \lambda_l \Theta(l-z), \label{sm:IC-1} \\
% \phi(z,0)&=&0 \implies \gamma(z,0)=\omega_{in}(z) \label{sm:IC-1} \\ 
%   h(z,T) &=& \lambda_k \Theta(k-z) + \lambda_l \Theta(l-z) \label{sm:IC-2}
%  \end{eqnarray}
and the following discontinuity conditions
\begin{eqnarray}
\left( \phi(z,t)~\partial_z\hst(z,t)~\right)_{0^+}^{0^-} &=& 0, \label{sm:DC-1} \\
\left(\frac{\mu_1}{2} \psi(z,t)~\partial_z \omegast (z,t) ~\right)_{0^+}^{0^-} &=& 
\left(\sigma(\omegast ) \psi(z,t) \partial_z\hst(z,t) ~\right)_{0^+}^{0^-}, \label{sm:DC-2}
\end{eqnarray}
Now we know that 
\begin{equation}
\boxed {q ~\omegast(z,t)|_{z \to 0^-}= p ~\omegast(z,t)|_{z \to 0^+}}, \label{sm:density-dis}
\end{equation}
which implies 
\begin{equation}
q~ \phi(z,t)|_{z \to 0^-}= p ~\phi(z,t)|_{z \to 0^+}. \label{sm:phi-dis}
\end{equation}
Using this in \eqref{sm:DC-1} we get 
\begin{equation}
p~\partial_z\hst(z,t)|_{z \to 0^-}= q ~\partial_z\hst(z,t)|_{z \to 0^+} \implies \boxed {\left(\omegast(z,t) \partial_z \hst(z,t) \right)_{0^+}^{0^-}=0}. \label{sm:DC-3} 
\end{equation}
We now need to find the discontinuities in the derivative of $\omegast (z,t)$ across $z=0$, for this we note from \eqref{H-continuity} that $\psi(z,t)|_{z \to 0^+}= \psi(z,t)|_{z \to 0^-}$. Using this in \eqref{sm:DC-2} we get 
\begin{equation}
\boxed{ \frac{\mu_1}{2}\left[\partial_z \omegast  \right]_{0^+}^{0^-}=  \left[\sigma(\omegast )\partial_z \hst\right]_{0^+}^{0^-}}. \label{sm:DC-4}
\end{equation}
This condition can also be obtained from the fact that the 
optimal current
%$j_{opt}(z,t) = $
$-\frac{\mu_1}{2} \partial_z \omegast  + \sigma(\omegast ) \partial_z \hst$ that is read from~\eqref{sm:opt-1}
is continuous across $z=0$ implying \eqref{sm:DC-4}.  
Using this in \eqref{sm:DC-2} we consistently get back 
$\psi(z,t)|_{z \to 0^-}=  \psi(z,t)|_{z \to 0^+}$. So finally we have to solve the optimal equations \eqref{sm:opt-1} and \eqref{sm:opt-2} with the following discontinuity conditions
\begin{eqnarray}
&&\hst(z,t)|_{0^+}^{0^-}= 0, \nonumber \\
&&q ~\omegast(z,t)|_{z\to 0^-}=p~\omegast(z,t)|_{z\to 0^+}~, \nonumber \\
&&\frac{\mu_1}{2}\left[\partial_z \omegast \right]_{0^+}^{0^-}=
\left[\sigma(\omegast)(z,t)\partial_z \hst(z,t)\right]_{0^+}^{0^-},\nonumber\\
&&\left[\omegast(z,t)\partial_z \hst(z,t)\right]_{0^+}^{0^-}=0\label{discontinuities} 
\end{eqnarray}
%where $A(z)|_{0^+}^{0^-}=A(0^-)-A(0^+)$ 
and the initial/boundary conditions :
\begin{eqnarray}
&&\omegast(z,0) = \omegain(z), \nonumber \\ 
%  (z,0)&=& - \lambda_k \Theta(k-z)- \lambda_l \redw{\Theta}(l-z) \label{BCS} \\
&& \hst(z,T)= \lambda_k \Theta(k-z)+\lambda_l \Theta(l-z), \label{f-BCs} \\
&& \omegast(z,t)|_{|z|\to \infty}= \omegatyp(z,t)_{|z|\to \infty}. \nonumber
\end{eqnarray}
where $\Theta(x)$ is the Heaviside theta function.

\section{Derivation of the correlation function}

%For a fixed arbitrary initial condition $\omegain(z)$ and large times, the generating function of the positions of particles $k$ and $l$ as written in~\eqref{sm:Path-int} is given by $\exp(-S[\omegast[\omegain] ,\hst[\omegain]]~)$ with the action~\eqref{sm:action} 
%evaluated at the optimal trajectories $\omegast $ and $\hst$  determined up to order $\lambda$ as functional of initial configuration $\omegain(z)$ \greenw{(see eq. in main text)}. 
%The optimal trajectories depend on the initial condition since the $\lambda^0$ order of $\omegast $ 
%is given by~\eqref{omegazanyic} and because at order $\lambda^1$ the mobility $\sigma(\omegatyp(z,t))$ appears in their expressions, with $\omegatyp(z,t)$ again given by~\eqref{omegazanyic}.
\noindent
After having computed the optimal action  (see \eqref{letter-action-1} of the main text) for a given initial condition, what remains is to average over the initial condition $\omegain$. This is done again using the saddle point method.

\subsection{Fixed initial condition}
%For the quenched uniform initial condition we do not need to perform the final average over initial configurations. This is equivalent to taking $\Sigma \to \infty$ in~\eqref{sm:var-annealed}, so that only the first term survives. We get
\noindent
For the fixed initial condition no integration over $\omegain$ is required.
%\greenw{See what to do what the derivative of $h$} \greenw{Until here}
%\begin{eqnarray}
%\tilde{S}[\omegain] &=&  -  \int_{-\infty}^{\infty} du \left( \lambda_k \int_{-\infty}^k dv~ (F^a(v,u,T)-\delta (v-u))~ + \lambda_l \int_{-\infty}^l dv~ (F^a(v,u,T)-\delta (v-u)) \right) \omegain(u) \nonumber \\ && 
%+ \frac{\mu_1 \mu_2}{2(\mu_1 - \mu_2)} \int_{-\infty}^{\infty} d u \int_{-\infty}^{\infty} d v \left( \int_{t=0}^T dt \int_{-\infty}^{\infty} d w (\partial_w h(w,t))^2 F^a(w,u,t) 
%F^a(w,v,t) \right) \omegain(u) \omegain(v).  \label{sm:action-omegai}  
%\end{eqnarray}
After expanding the action up to order $\mathcal{O}(\lambda^2)$ and using the expression of $\omegatyp$ from \eqref{letter-omega_0} of the main text, one can get the correlation $c^{fix}_{k,l}=\langle X_k(T) X_l(T) \rangle^{fix}_c$ from the coefficient of $\lambda_k\lambda_l$ as 
\begin{eqnarray}
c^{fix}_{k,l}[\omega_{in}]  &=& \frac{\mu_1 \mu_2}{2(\mu_1 - \mu_2)} \int_{t=0}^T dt \int_{-\infty}^{\infty} dz~ \partial_z h_k(z,t) \partial_z h_l(z,t)~\omegatyp(z,t)^2  ~~~\label{c-fix} \nonumber \\
&=& \frac{\mu_1 \mu_2}{2(\mu_1 - \mu_2)} \int_{t=0}^T dt \int_{-\infty}^{\infty} dz~ \partial_z h_k(z,t) \partial_z h_l(z,t)
\\ && \times \left(\int_{-\infty}^{\infty} d u  \nonumber F_b(z,u,t) \omega_{in}(u) \right) 
\left(\int_{-\infty}^{\infty} d v F_b(z,v,t) \omega_{in}(v)\right), \nonumber
\end{eqnarray}
where $F_b(z,z',t)$ and $h_\epsilon(z,t)$ are given in \eqref{letter-Fa} and~\eqref{letter-sol-h_k} of the main text.
\subsection{Random initial condition}
%After we obtain the optimal value of the action $S[\omegast[\omegain] ,\hst[\omegain]]$ for given initial configuration $\omegain$, we now have to perform the average over the initial configurations as in equation~\eqref{letter-Path-int} in the main text. 
\noindent
We now chose initial configurations from the distribution $\mathcal{P}_{in}[\omegain]=e^{-G[\omegain]}$.
From equations \eqref{letter-cumu-func} and \eqref{letter-Path-int} of the main text, 
the generating function can be written as 
\begin{equation}
\langle \exp[\lambda_kX_k(T)+\lambda_lX_l(T)] \rangle_{\mathcal{P}_{in}} = \int \mathcal{D}\omegain~e^{-(G[\omegain]+S[\omegain])}, \label{sm:Path-int-omegai}
\end{equation}
where the 'new' action $\tilde{S}[\omegain] = G[\omegain] +S[\omegast[\omegain],\hst[\omegain]]$ is given by 
\begin{eqnarray}
%\tilde{S}[\omegain] &=& G[\omegain] +S[\omegast[\omegain],\hst[\omegain]] \nonumber \\
\tilde{S}[\omegain]&=& G[\omegain]- \int_{-\infty}^{\infty} du \bigg( \lambda_k \int_{-\infty}^k dv~ (F_b(v,u,T)-\delta (v-u))~ \nonumber \\ &&~~~~~~~~~+ \lambda_l \int_{-\infty}^l dv~ (F_b(v,u,T)-\delta (v-u)) \bigg) \omegain(u) \label{sm:action-omegai} \\ && \hspace{-15mm}
+ \frac{\mu_1 \mu_2}{2(\mu_1 - \mu_2)} \int_{-\infty}^{\infty} d u \int_{-\infty}^{\infty} d v \left( \int_{t=0}^T dt \int_{-\infty}^{\infty} d z (\partial_z h(z,t))^2 F_b(z,u,t) 
F_b(z,v,t) \right) \omegain(u) \omegain(v).   \nonumber  
\end{eqnarray}
To optimize the action $\tilde{S}$, $\omega_{in}(z)$ must satisfy $\frac{\delta \tilde{S}}{\delta  \omega_{in}} = 0$.
The functional $G[\omega(z)]$ usually has the following expansion around the most probable profile $\chi(z)$:
\begin{equation}
G[\omega] = \int du ~\left( \frac{(\omega-\chi)^2}{2 \Sigma[\chi(u)]^2} +\mathcal{O}((\omega-\chi)^3)+ .... \right)
%, ~~~\text{where}~~\Sigma[\chi(u)]=(\delta^2 G /\delta \omega^2)_{\omega(u)=\chi(u)}
.\nonumber 
\end{equation}
Using this in \eqref{sm:action-omegai} we get 
an equation for the optimal solution $\chist$,
\begin{eqnarray}
&&\chist(z) + \Sigma^2(z) \frac{ \mu_1 \mu_2}{\mu_1 - \mu_2}  \int_{-\infty}^{\infty} d v \left(  \int_{t=0}^T dt \int_{-\infty}^{\infty} du~ (\partial_u h(u,t))^2 F_b(u,z,t) F_b(u,v,t) \right) 
\chist(v) \nonumber \\
&=& \chi(z) + \Sigma^2(z) \left( \lambda_k  \int_{-\infty}^k dv  (F_b(z,v,T)-\delta (z-v)) + \lambda_l  \int_{-\infty}^l dv (F_b(z,v,T)-\delta (z-v)) \right) \label{sm:deltai-opt}  
\end{eqnarray}
Eq.~\eqref{sm:deltai-opt} is in general a complicated integral equation, but the integral kernel on the LHS is of order $\lambda^2$, so that the solution can be expanded in $\lambda$. Moreover, 
we will see that knowing $\chist$ to order $\lambda$ is enough to obtain $\tilde{S}$ to the desired order $\lambda^2$. Straightforward manipulations give
\begin{equation}
\chist(z) = \chi(z) + \Sigma^2(z) \left( \lambda_k \mathcal{F}^b_k(z,T) + \lambda_l \mathcal{F}^b_l(z,T) \right) + \mathcal{O}(\lambda^2), \label{sm:solchi}
\end{equation}
where we have defined
\begin{eqnarray}
\mathcal{F}^b_\epsilon(u,t) &=& \int_{v=-\infty}^\epsilon (F^b(v,u,t)-\delta (u-v)) dv 
=\frac{1}{2}\left(\text{erfc}\left(\frac{|u-\epsilon|}{\sqrt{2 \mu_1  t}}\right) + b~ \text{erfc}\left(\frac{\left| \epsilon\right| +\left| u\right| }{\sqrt{2 \mu_1  t}}\right)\right).
\label{defcF}
\end{eqnarray}
The optimal action reads
\begin{eqnarray}
\tilde{S}[\chist] &=& \int_{-\infty}^\infty du~ \frac{(\chist(u) - \chi(u))^2}{2 \Sigma(u)^2} \nonumber \\ && - \int_{-\infty}^{\infty} du \left( \lambda_k \int_{-\infty}^k dv ~ (F_b(v,u,T)-\delta (v-u)) + \lambda_l \int_{-\infty}^l dv ~ (F_b(v,u,T)-\delta (v-u)) \right) \chist(u)  \\ 
&&+ \frac{\mu_1 \mu_2}{2(\mu_1 - \mu_2)} \int_{-\infty}^{\infty} d u \int_{-\infty}^{\infty} d v \left( \int_{t=0}^T dt \int_{-\infty}^{\infty} dz~ (\partial_z \hst(z,t))^2 F_b(z,u,t) 
F_b(z,v,t) \right) \chist(u) \chist(v) \nonumber \\
&=& - \int_{-\infty}^\infty du \Sigma(u)^2 \frac{(\lambda_k  \mathcal{F}^b_k(u,T) + \lambda_l  \mathcal{F}^b_l(u,T))^2}{2} - \int_{-\infty}^{\infty} du \left( \lambda_k \mathcal{F}^b_k(u,T) + \lambda_l \mathcal{F}^b_l(u,T) \right) \chi(u) \\
&&+ \frac{\mu_1 \mu_2}{2(\mu_1 - \mu_2)} \int_{-\infty}^{\infty} d u \int_{-\infty}^{\infty} d v \left( \int_{t=0}^T dt \int_{-\infty}^{\infty} dz~ (\partial_z \hst(z,t))^2 F_b(z,u,t) 
F_b(z,v,t) \right) \chi(u) \chi(v) \nonumber \\ &&+ \mathcal{O}(\lambda^3). \nonumber  \label{sm:action-chii}  
\end{eqnarray}
Note that $\hst(w,t)$ is already of $\mathcal{O}(\lambda)$ as we see from \eqref{letter-expnsn-2} of main text.
From the $\mathcal{O}(\lambda)$ terms of the above equation we get $\langle X_\epsilon(T) \rangle = \int_{u=-\infty}^\infty \mathcal{F}^b_\epsilon(u,T) \chi(u) du$, which matches the result in \eqref{letter-mcal-Y} from the main text when $\chi(z) = \nu_0$. From the $\lambda_k \lambda_l$ terms of $\tilde{S}[\chist]$ we get $c^{ran}_{k,l}=\langle X_k(T) X_l(T) \rangle^{ran}_c$ as
\begin{eqnarray}
c^{ran}_{k,l}[\chi]&=&c^{fix}_{k,l}[\chi]  +  \int_{-\infty}^\infty du \Sigma(u)^2 \mathcal{F}^b_k(u,T) \mathcal{F}^b_l(u,T), \label{c-rand}
\end{eqnarray}
where $\mathcal{F}^b_\epsilon(z,T)$ is given in 
\eqref{defcF}.

\section{Variance of the tracer position for arbitrary bias}

\subsection{Quenched uniform initial condition}

Starting from \eqref{c-fix} and putting $k=0,~l=0$ and $\omegain=\nu_0$ we compute
\begin{eqnarray}
c^{QU}_{0,0}  &=& \frac{\mu_1 \mu_2}{2(\mu_1 - \mu_2)} \nu_0^2  \int_{t=0}^T dt \int_{-\infty}^{\infty} dz~ \left(\partial_z h_0(z,t) \int_{u=-\infty}^\infty du~ F_b(z,u,t) \right)^2 \nonumber \\
&=& \frac{\mu_1 \mu_2}{2\pi (\mu_1 - \mu_2)} \nu_0^2  \int_{t=0}^T dt \int_{-\infty}^{\infty} dw~ \left(\frac{1+b ~\text{sgn}(w)}{\sqrt{2 \mu_1 (T-t)} } \right)^2 e^{-\frac{w^2}{ \mu_1 (T-t)}}\left( 1-b~ \text{sgn}(w)~ \text{erfc}\left(\frac{|w|}{\sqrt{2 \mu_1  t}}\right)\right)^2 \nonumber \\
&=& \frac{\mu_1 \mu_2}{2 \pi (\mu_1 - \mu_2)} \nu_0^2 \frac{\sqrt{T}}{\sqrt{2 \mu_1}} \int_{\tau=0}^1 d\tau \int_{-\infty}^{\infty} dw~ \frac{1}{1-\tau}\left(1+b ~\text{sgn}(w) \right)^2 e^{-\frac{2 w^2}{ 1-\tau}}\left( 1-b~ \text{sgn}(w)~ \text{erfc}\left(\frac{|w|}{\sqrt{\tau}}\right)\right)^2, \nonumber \\
\label{Aqu}
\end{eqnarray}
where in the second line we used explicit expressions of $h_0(z,t)$ and $F_b(z,z',t)$ from eqs. \eqref{letter-sol-h_k} and \eqref{letter-Fa} of the main text.
Now using the following integral identities, 
\begin{eqnarray}
4 \int_{s=0}^1 ds \int_{y=0}^\infty dy ~\text{erf}^n \left( \frac{y}{\sqrt{s}}\right) e^{-\frac{2 y^2}{1-s}} &=& 2 \sqrt{2 \pi} (\sqrt{2}-1)^n~\text{for}~n=0,1,2,
\end{eqnarray}
we obtain
\begin{equation}
c_{0,0}^\text{QU}(T)=\nu_0^2\sqrt{2 \mu_1 T}~\frac{\mu_2}{\sqrt{\pi}(\mu_1-\mu_2)}~\sqrt{2} (\sqrt{2}-1)^2 ~ \left[ (\sqrt{2}+1) (1-b^2) +\frac{1}{2} (1+b^2)\right].
\end{equation}

\subsection{Quenched linear initial condition}
\noindent
For this case $\omegain(z') =\alpha |z'|$. Inserting this in \eqref{letter-omega_0} of the main text and performing the integral we get 
\begin{equation}
\omegatyp(z,t) = \sqrt{2\mu_1 t}~\alpha~\Omega_b^{QL}\left(\frac{z}{\sqrt{2\mu_1t}}\right),~~\text{where}~~\Omega_b^{QL}(u)=|u|+[1-b~\text{sgn}(u)][{e^{-u^2}}/{\sqrt{\pi}}-|u|\text{Erfc}(|u|)]. \label{O-typ-ql}
\end{equation}

Inserting this result for $\omegatyp(z,t)$  in \eqref{c-fix} and simplifying, we get 
\begin{equation}
c_{k,k}^{QL}(T)= \alpha^2 T^{3/2}~ \mathcal{C}_b^{QL}\left(\frac{|k|}{\sqrt{2\mu_1T}},\frac{|k|}{\sqrt{2\mu_1T}}\right),
\end{equation}
where,
\begin{eqnarray}
\mathcal{C}_b^{QL}(u,u)&=&\frac{\sqrt{2\mu_1}\mu_1 \mu_2}{2 \pi(\mu_1 - \mu_2)} \int_{\tau=0}^1 d\tau \frac{\tau}{1-\tau} \int_{-\infty}^{\infty} dz'~\left(e^{-\frac{(z'-u)^2}{1-\tau}}+b~\text{sgn}(z')~e^{-\frac{(|z'|+|u|)^2}{1-\tau}}\right)^2 \nonumber \\ 
&& ~~~~~~~~~~\times~\left( \frac{|z'|}{\sqrt{\tau}}+[1-b~\text{sgn}(z')]\left[\frac{e^{-\frac{z'^2}{\tau}}}{\sqrt{\pi}}-\frac{|z'|}{\sqrt{\tau}}\text{Erfc}\left(\frac{|z'|}{\sqrt{\tau}}\right)\right]\right)^2 . 
\end{eqnarray}

\subsection{Steady state initial condition}
Starting from \eqref{c-rand} and putting $k=0,~l=0$ and $\langle \omegain \rangle_{\mathcal{P}_{ss}}=\nu_0$ we get
\begin{eqnarray}
c^{SS}_{0,0}  - c^{QU}_{0,0}  &=& \frac{\mu_2}{\mu_1-\mu_2} \nu_0^2\int_{-\infty}^\infty du \mathcal{F}^b_0(u,T)^2 \nonumber \\
&=& \frac{ \mu_2}{4(\mu_1-\mu_2)} \nu_0^2 \sqrt{2 \mu_1 T} (1+b^2) \int_{-\infty}^\infty dw ~ \text{erfc}\left(\left| w\right|\right)^2 \\
&=& \nu_0^2 \sqrt{2 \mu_1 T}  \frac{ \mu_2}{\sqrt{\pi}(\mu_1-\mu_2)} \frac{\sqrt{2}-1}{\sqrt{2}} (1+b^2),\nonumber
\label{Ass}
\end{eqnarray}
hence the variance of the tracer position,
\begin{equation}
c_{0,0}^\text{SS}(T)=\nu_0^2\sqrt{2 \mu_1 T}~\frac{\mu_2}{\sqrt{\pi}(\mu_1-\mu_2)}~\sqrt{2} (\sqrt{2}-1)^2 ~ \left[ (\sqrt{2}+1) (1-b^2) +\frac{2+\sqrt{2}}{2} (1+b^2)\right].
\end{equation}
\end{onecolumn}

\end{document}